# Discrete breathers of new type in monoatomic chains

G.M. Chechin[†] and V.S. Lapina

[†]e-mail: gchechin@gmail.com

Southern Federal University, Institute of physics,
Stachki Ave., 194, 344090, Rostov-on-Don, Russia

In strained monoatomic chains with Lennard-Jones interactions, we revealed a stable static non-homogeneous structure appearing as a result of a certain phase transition. Positions of individual particles in this structure form an exact arithmetic progression whose difference depends on the value of the strain. For N-particle chain, this structure is characterized by one long and N-1 short interatomic distances (bonds). In the vicinity of the static structure, we found discrete breathers of new type which essentially differ from the traditional breathers in the form of Sievers-Takeno and Page modes. It is well known that these modes possess some staggered structures and demonstrate exponential decay of the particle amplitudes from the core to their tails. In contrast to such properties, our breathers are characterised by smooth decay and amplitudes of the particles form approximately a decreasing arithmetic progression. Core of these breathers is located on two particles with long bond in static structure. Our breathers demonstrate soft type of nonlinearity (the frequency decreases with increasing of amplitudes) and they are stable dynamical objects for amplitudes up to 20%-30% of interparticle distance of the strained equidistant chain. For infinitely small amplitudes these breathers tend to the above described static non-homogeneous structure. We studied dependence of their properties on amplitude, strain and the number of particles in the chain. There exist a reason to suppose that the above static and dynamical structures can exist in real monoatomic chains consisting of carbon, boron, and other atoms.

В растянутых одноатомных цепочках с взаимодействиями Леннарда-Джонса мы обнаружили устойчивую неоднородную статическую структуру, возникающую в результате некоторого фазового перехода. Положения отдельных частиц в этой структуре образуют точную арифметическую прогрессию, разность которой зависит от величины растяжения. Для N-частичной цепочки эта структура характеризуется одним длинным и N-1 короткими межатомными расстояниями (связями). Вблизи найденной статической структуры мы обнаружили дискретные бризеры нового типа, которые существенно отличаются от традиционных бризеров в форме мод Сиверса-Такено и Пейджа. Хорошо известно, что эти моды обладают некоторыми ступенчатыми структурами и демонстрируют экспоненциальное убывание амплитуд частиц от их ядра к хвостам. В отличие от этих свойств, наши бризеры характеризуются плавным профилем, а амплитуды частиц приближенно образуют убывающую арифметическую прогрессию. Ядро этих бризеров расположено на двух частицах с длинной связью в статической структуре. Наши бризеры демонстрируют мягкий тип нелинейности (частота уменьшается с увеличением амплитуды) и являются стабильными динамическими объектами для амплитуд вплоть до 20% -30% от межчастичного расстояния растянутой эквидистантной цепочки. Для бесконечно малых амплитуд эти бризеры стремятся к описанной выше статической неоднородной структуре. Нами была изучена зависимость их свойств от амплитуды, деформации и количества частиц в цепочке. Есть основания полагать, что вышеуказанные статические и динамические структуры могут существовать в реальных моноатомных цепочках, состоящих из атомов углерода, бора и других элементов.

## 1. Introduction

The study of discrete breathers (DB) in systems of differents dimensions, structure and various physical nature is an actual problem of modern condensed matter physics (see the review papers by Aubry [1], Flach and Gorbach [2], Dmitriev [3]), since these dynamical objects can significantly affect the physical properties of the corresponding materials. The review paper by Flach-Gorbach [4] is devoted to the consideration of discrete breathers in the form of Sievers-Takeno (symmetric DB) and Page modes (antisymmetric DB) in the monoatomic chains without on-site potential.

In the present work, DBs of an essentially different type were found in the strained monoatomic chain with the Lennard-Jones interparticle potential. Such type of discrete breathers can exist in the monoatomic chains with other interparticle potentials. Their comparison with "traditional" discrete breathers will be discussed in Conclusion.

## 2. Mathematical model

We consider equidistant monoatomic chains whose interparticle interaction is described by the Lennard-Jones (L-J) potential

$$\varphi(r) = \frac{A}{r^{12}} - \frac{B}{r^6}. \tag{1}$$

In this case, the interaction is considered only between nearest neighbours and periodic boundary conditions is assumed. Using the appropriate scaling of the time and space variables in the dynamical equations of the considered chain, induced by the potential (1), one can obtain $A = B = 1$ without loss of generality.

Then interaction force between two particles at a distance r from each other is determined by the formula:

$$f(r) = -\frac{d\varphi}{dr} = \frac{12}{r^{13}} - \frac{6}{r^7}. \tag{2}$$

The equilibrium distance between these particles $a_0$, determined from the condition that their interaction force is zero, is $a_0 = \sqrt[6]{2} \approx 1.12246$.

This is the equilibrium interparticle distance in the *unstrained* L=J chain ($\eta = 0\%$). Under the strain $\eta \neq 0\%$, the interparticle distance $a$ increases:

$$a = a_0(1+\eta). \tag{3}$$

We denote by $x_i (i = 1..N)$ the *longitudinal* displacements of particles from their equilibrium positions in the *equidistant* structure of the monoatomic chain with an arbitrary pair interparticle potential $\varphi(r)$ and periodic boundary conditions ($x_{N+1} = x_1$). Then the potential energy of the chain $U(x_1, x_2, ..., x_N)$ can be written in the form:

$$U(x_1, x_2, \ldots, x_N) = \sum_1^N \varphi(x_{i+1} - x_i). \tag{4}$$

The equilibrium state of the chain corresponds to a minimum of the potential energy, so that it satisfies the condition

$$\frac{\partial U}{\partial x_i} = 0, i = 1..N. \tag{5}$$

## 3. Static structure of strained Lennard-Jones chains

The equidistant structure is not the only possible static structure of the monoatomic chains. Let us consider this issue in more detail.

In the absence of the strain, the equidistant structure of the chain under consideration takes place: all the interatomic distances are the same and equal to $a_0 = 1.12246$. The symmetry group of such a chain (the periodic boundary conditions are assumed) contains $N$ translations by multiples of the period $a_0$ of our one-dimensional lattice and N inversions centered at the interatomic distances (symmetry group $G_0 = D_N$). However, in the presence of a strain, the situation can change radically! In papers [5-7], Peierls phase transition is considered, which is connected with doubling of the lattice period. As a result of such a transition, all the odd translations disappear and the symmetry group decreases by half (we mean the even values of N). It represents one of the subgroups of the symmetry group $G_0$.

This is only an example of the general statements of the phase transitions Landau theory: structural transitions in crystals are associated with a spontaneous lowering of the symmetry under the change of external parameters such as temperature or pressure. Let us emphasize that these parameters are *scalar* and, at first glance, their change can not lead to a decrease of the crystal symmetry. However, for some values of these parameters, the crystal lattice becomes *unstable* and its symmetry decreases *spontaneously*. In the one-dimensional case under consideration, the strain of the chain is analogous to applying negative pressure to the crystal. It will be shown below, that this strain can lead, at its appropriate values, to a phase transition associated with the loss of all translations and inversions, except one that passes through the "center" of the chain (because of periodic boundary conditions our chain can be represented as a ring and such a center can be chosen at any site).

However, the static structure of the monoatomic chain, that we revealed, can not be explained only by the symmetry lowering and seems to be nontrivial. Indeed, the set of atomic displacements from its "legal" positions in the original *equidistant* structure form some *exact (!) arithmetic progression*.

Thus, we consider the following static structure of a chain with an arbitrary pair interaction potential $\varphi(r)$:

$$\{0, q, 2q, 3q, ..., (m-1)q, mq \mid -mq, -(m-1)q, ..., -3q, -2q, -q, 0\}. \tag{6}$$

Here $2m + 1 = N$ is the number of particles in the chain. The center of the chain is marked by a vertical dash (the inversion passes through it). Thus, the left half of the particles have positive displacement values, which form an arithmetic progression with difference q, and the right half of the particles have exactly the same displacements in magnitude but opposite in sign (on the right, at the end of Eq. 6, a particle with zero displacement is added due to periodic boundary conditions).

For clarity, in Fig. 1 we show our chain in the form of the ring for the case $N = 7$. The points correspond to the equidistant structure, and the displacements of the particles corresponding to the structure (6) are represented by arrows.

For the artificially introduced static structure (6) and for an arbitrary potential $\varphi(r)$, it is easy to find the total potential energy, noting that there are $N-1$ short interatomic bonds (a-q) and only one long bond $[a + (N-1)q]$. In this way, we obtain

$$U(q) = (N-1)\varphi(a-q) + \varphi[a+(N-1)q], \tag{7}$$

where $a = a_0(1 + \eta)$ is the interatomic distance in the equidistant structure with $\eta\%$ strain. From the condition of minimum of the energy (7), we find an equation, which we call hereafter "force equation":

$$f(a-q) = f[a+(N-1)q]. \tag{8}$$

Here, $f(r)$ is the force corresponding to the pair potential $\varphi(r)$ [for the L-J potential it is given by Eq. (2)]. Actually equation (8) means the equality of forces acting on the maximally displaced particle (in Fig. 1 it has number 4) from the left neighbour (short bond) and the right neighbour (long bond).

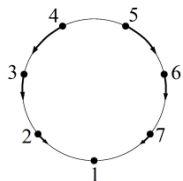

**Fig.1.** Ring diagram of the monoatomic chain with the static structure (6) for $N=7$ particles.
**Рис.1.** Круговая диаграмма моноатомной цепочки со статической структурой (6) для случая N=7 частиц.

The nonlinear algebraic equation (8) has several real roots, which give possible values of the difference q of the arithmetic progression. In the case of the L-J potential, these roots can be easily found using the mathematical package MAPLE. In this paper, we analyze two closest to zero roots of this equation by graphical method.

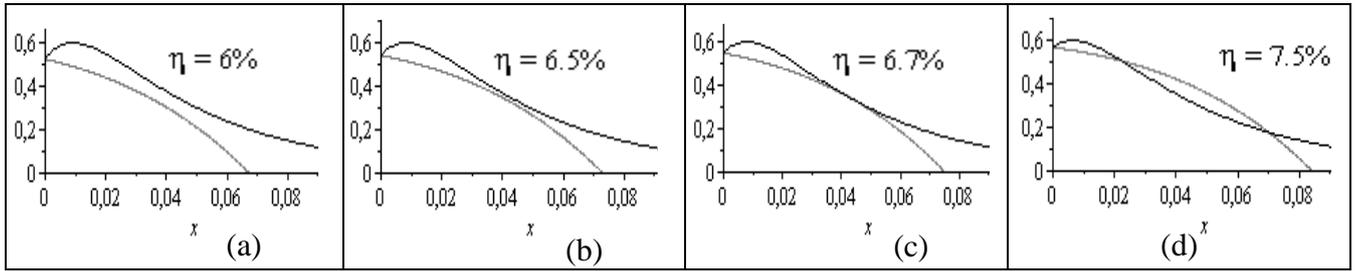

**Fig. 2.** Graphical analysis of the roots of the force equation (8) for N=7.
**Рис. 2.** Графический анализ корней силового уравнения (8).

Figs.2a-2d show the plots of the left side of the force equation (black colour) and the right part (gray colour) for N = 7 and the strain range [6%; 7.5%]. Obviously, the *intersection* points of these plots give the possible values of the arithmetic progression difference q. All these plots have one common point q = 0, which corresponds to the trivial solution of equidistant chain structure. Fig. 2a and 2b show that for the tension of 6% and 6.5%, the point q = 0 is the only one point of intersection of the left and right parts of the force equation (8), although there is a tendency for them to approach each other outside the origin. In Fig. 2c, corresponding to a tension of 6.7%, one can see the appearance of the point of contact of the plots under consideration. It determines the *bifurcation* of the solution of the force equation, after which two new roots of this equation appear, which is clearly seen in Fig. 2d. It occurs that the larger root ($x_1$) corresponds to the stable non-homogeneous static structure of the chain, while the second root ($x_2$) corresponds to its unstable structure.

The roots $x_1$ and $x_2$ are removed from each other with increasing strain of the chain. In this case, the root $x_2$ moves toward the origin and vanishes at 11% of the strain. This root is of no interest to us, since it generates an *unstable* static structure. On the other hand, the root $x_1$ generates a *stable* non-homogeneous static structure of the L-J chain, in the vicinity of which there exist discrete breathers of a new type.

The discussed pictures correspond to the case N = 7. It is interesting to consider the behaviour of the root $x_1$ with increasing number of particles in the chain (N). The results of this study are presented in Table 1 for the case of 10% strain. It is clear from this table that the root $x_1$ increases slowly with increasing N and tends to some limit. It is interesting that this limit can be found analytically. Indeed, let us consider the force equation (8) for N tending to infinity. The argument of the force on the right-hand side in this equation also tends to infinity. Therefore, this force, with an infinite distance between the interacting particles, tends to zero. So, from the left side of the force equation we have *f (a - q) = 0*, where a = $a_0$(1 +η) and $a_0$ = 1.12246. However, the L-J force in the finite interval of its argument is equal to zero only at the minimum potential point, i.e. a + q = $a_0$. Then we have $a_0$(1 +η)-q=$a_0$ and, consequently, q = $a_0$η. The data in Table 1 are given for the strain η= 10% and, therefore, q = 1.12246*1/10 = 0.112246. This value of the difference q of the arithmetic progression coincides with the limit value of this difference given in Table 1.

**Таблица 1.** Зависимость корня $q_1$ силового уравнения (8) от числа частиц в цепочке N.
**Table 1**. Dependence of the root $q_1$ of the force equation (8) on the number N particles in the chain

| N | 7 | 9 | 11 | 13 | 15 | 17 | 19 | 21 | 23 | 25 | 27 | 29 |

| $q_1$ | 0.1071 | 0.1100 | 0.1100 | 0.1116 | 0.1119 | 0.1120 | 0.1121 | 0.1121 | 0.1122 | 0.1122 | 0.1122 | 0.1122 |

The critical value of the strain $\eta_c$, at which the non-zero root $x_1 = q$ appears for the first time, depends on the number N of particles of the chain. Table 2 shows the dependence $\eta_c$ (N).

**Таблица 2.** Зависимость $\eta_c$ (N).
**Table 2**. Dependence $\eta_c$ (N).

| N | 5 | 7 | 11 | 13 | 15 | 19 | 23 | 27 | 33 | 43 | 53 | 103 |
|---|---|---|---|---|---|---|---|---|---|---|---|---|
| $\eta_c$ | 8.05% | 6.71% | 5.10% | 4.65% | 4.20% | 3.70% | 3.25% | 2.89% | 2.60% | 2.20% | 2.00% | 1.25% |

### 4. Discrete breathers of new type

To construct a discrete breather in the neighbourhood of the static structure discussed in the previous section, it is necessary to determine its initial profile, which in turn, determines the initial conditions for solving Newton's differential equations describing the dynamics of the L-J chain. These initial conditions are formed by a set of displacements of all particles from their equilibrium positions specified by the static structure, under the assumption that their initial velocities are zero. The required displacements from the static structure (6) will be given in the form of a numerical sequence whose terms are proportional to the structure (6) itself with some amplitude A. Thus, we take the initial profile of the breather in the form:

$$\{(1 + A) kq\} \mid k = 0..N\}, \qquad (9)$$

where $q$ is the difference of the arithmetic progression that determines our static structure. Why in this form? The point is that we found accidentally quasi-breathers of the type (9), and only after that we began to study the static structure in the neighbourhood of which these dynamical objects can exist.

In Figs. 3a-3e, for the case N = 7 and $\eta$ = 10%, the time evolution of dynamical objects, determined by the initial conditions (9), is presented. They are used for solving Newton's differential equations describing dynamics of our chain that depends on their amplitude A.

For A = 0, straight lines are obtained for all particles, which correspond to the static structure (6). Let us remember that this structure determines the static displacements of the particles from their "legal" equilibrium positions in the strained L-J chain.

At small but non-zero amplitudes A, almost harmonic oscillations take place (Fig.3a), whereas for large amplitudes, the oscillations of the particles become essentially nonlinear. Figs. 3b-3e describe the time evolution of discrete breathers at amplitudes A = 0.1, 0.2, 0.3, and 0.4. These amplitudes are very significant, if one takes into account that the interparticle distance in the equilibrium L-J chain without strain is equal to 1.12246. However, already at A = 0.5, stability of the discrete breather is lost, and Fig. 3f shows the chaotic motion of the chain particles.

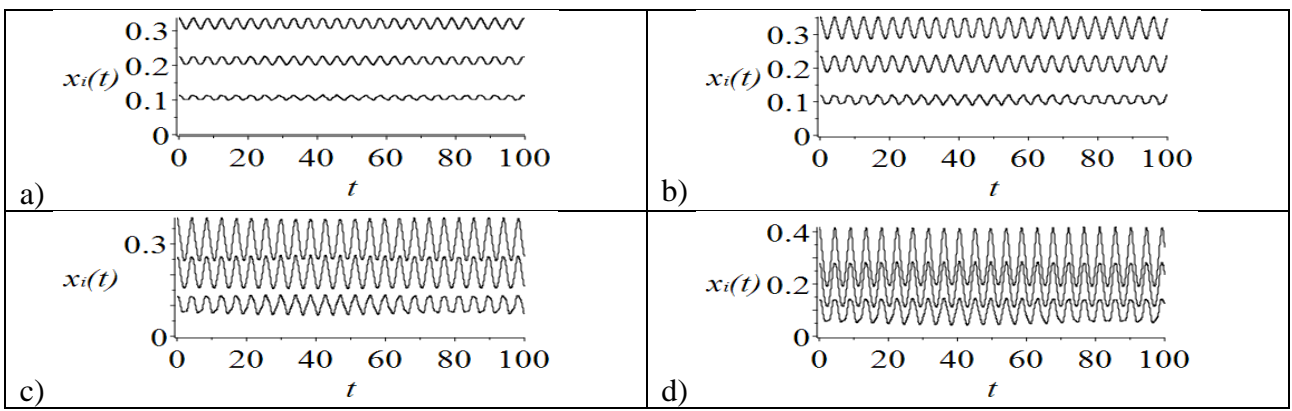

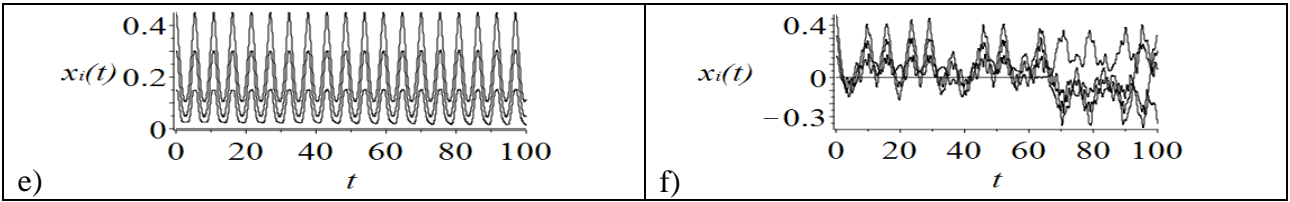

**Fig. 3.** Discrete breathers (quasi-breathers) of different amplitudes A in the Lennard-Jones chain with N=7 particles for the strain η=10%: A=0.05 (a), A=0.1 (b), A=0.2 (c), A=0.3 (d), A=0.4 (e), A=0.5 (f).

**Рис. 3.** Дискретные бризеры (квазибризеры) в цепочке Леннарда-Джонса для случая N=7 частиц при растяжении η=10% при амплитудах: A=0.05 (a), A=0.1 (b), A=0.2 (c), A=0.3 (d), A=0.4 (e), A=0.5 (f).

For completeness, Figs. S2a-S2d show the time evolution of quasi-breathers, which were constructed on the basis of the static structure (6) with deviations from it in opposite direction as compared to that in Figs. 3c-3f. In other words, we use for constructing these figures the initial profile of the form *{(1-A)kq | k = 0 .. (N-1)}*.

The dynamical objects presented in Figs. 3a-3e are not exact discrete breathers, but only quasi-breathers [8], because of the absence of ideal periodicity of oscillations. However, it seems that they are close enough to the exact breathers. In this paper, we did not aim to refine the profiles of the constructed quasi-breathers, as it was done in Ref. [9].

Figs. 4 and Fig. S3 show the quasi-breathers which we obtained for a large number of particles in the L-J chain: N = 19 (Figs. 4a, 4b) and N =27 (Figs. S3a, S3b) when the particles deviate *up* from the static structure with the amplitude A=0.3 and A = 0.4 (colour on-line). The time evolution of quasi-breather becomes more and more "expressive" with the increase in the number of particles N. Moreover, the amplitudes of oscillations of the particles from the breather core to its periphery demonstrate approximately a decreasing arithmetic progression. Note that in all figures we depict displacements only of the half of particles with positive values. The other half of particles possesses the same, but negative displacements.

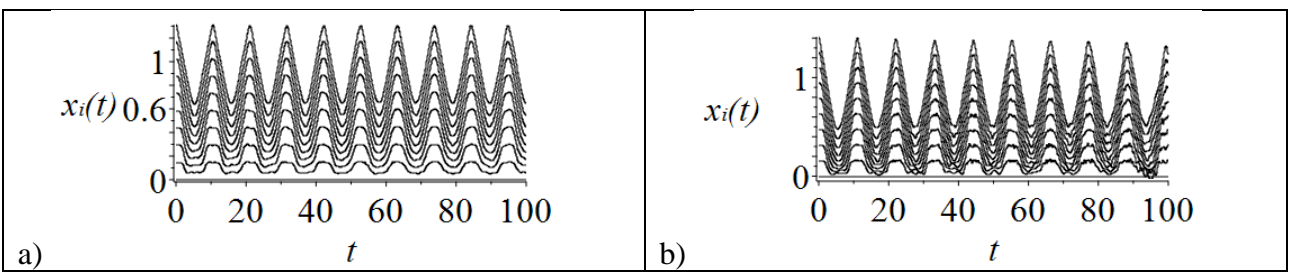

**Fig. 4.** Discrete breathers (quasi-breathers) of different amplitudes A in the Lennard-Jones chain with N=19 particles for the strain η=10%: A=0.3 (a), A=0.4 (b).

**Рис. 4.** Дискретные бризеры (квазибризеры) в цепочке Леннарда-Джонса для случая N=19 частиц при растяжении η=10% при амплитудах: A=0.3 (a), A=0.4 (b).

## 5. Conclusion

What is the difference between our discrete breathers (quasi-breathers) from traditional breathers in monoatomic chains? In the former dynamical objects, all particles oscillate near some static *equidistant* structure and they simultaneously take on zero values. This property is characteristic of both Sievers-Takeno and Page modes. Unlike this, in our breathers, the particles of the chain pass simultaneously through their *different* equilibrium positions, which form a *non-equidistant* structure, and this structure represents an *exact* arithmetic progression.

The instantaneous positions of the particles in traditional breathers form a *staggered* structure (the signs of the displacements of neighbouring particles alternate on passing from the core of the breather to its periphery). In our case, half of the neighbouring particles of the chain possess displacements of the same sign, and the other half possesses displacements of the opposite sign.

The traditional discrete breathers demonstrate usually an exponential decay of vibration amplitudes from the core of the breather to its periphery. Since the equilibrium positions of the traditional breather form a discrete equidistant structure, this type of decay means that the oscillation amplitudes of the particles form some *geometric* progression. In the case of our breathers, they form, as we know, some *arithmetical* progression. Fig. S3a shows this decay of the oscillation amplitudes from the core of the breather to its periphery for our breather in the case of N = 19, η = 10%. In fact, the set of these amplitudes forms two triangles above and below the horizontal axis.

Let us note that the displacements of two particles having maximum amplitudes (in absolute value) are similar to the core (... x, -x ...) of the Page mode, but the amplitudes of all other particles possess a completely different form. For comparison, we depict in Fig. S3b schematic representation of the Page mode with the maximal amplitude equal to that in Fig. S3a.

The presented analysis of the non-homogeneous static structure for Lennard-Jones chains is obviously also applicable to other pair potentials of interparticle interaction, for example, for Morse-type potentials, etc., since it is based only on the solution of the force equation (8), which is written for an arbitrary potential φ(r). Thus, we have proved that, for certain strains above the critical values, a stable static structure can be realized, exactly described by some arithmetic progression. The possibility of the formation of such structure and existence of discrete breathers in its vicinity we plan to study in the framework of DFT-models for the chains constructed from carbon and boron atoms.

**Acknowledgments**

*The authors acknowledge support by the Ministry of Education and Science of the Russian Federation (state assignment grant No. 3.5710.2017/8.9).*